\documentclass[aps,prd,preprint,tightenlines,superscriptaddress]{revtex4}
\usepackage{epsfig}
\usepackage{graphicx}
\usepackage{psfrag}
\usepackage{amsmath,amssymb}
\usepackage{colordvi}
\usepackage{color}
\usepackage{amsfonts}
\usepackage{enumerate}
\usepackage{slashed,bm}

\newcommand{\B}[1]{\textcolor{blue}{#1}}

\def \non {\nonumber}
\def \beq  {\begin{equation}}
\def \eeq  {\end{equation}}

\begin{document}

\title{Improved quasi parton distribution through Wilson line renormalization}
\author{Jiunn-Wei Chen}
\affiliation{Department of Physics, Center for Theoretical Sciences, and Leung Center for Cosmology and Particle Astrophysics, National Taiwan University, Taipei, Taiwan 106}
\affiliation{Center for Theoretical Physics, Massachusetts Institute of Technology, Cambridge, MA 02139, USA}
\author{Xiangdong Ji}
\affiliation{INPAC, Department of Physics and Astronomy, Shanghai Jiao Tong University, Shanghai, 200240, P. R. China}
\affiliation{Maryland Center for Fundamental Physics, Department of Physics,  University of Maryland, College Park, Maryland 20742, USA}
\author{Jian-Hui Zhang}
\affiliation{Institut f\"ur Theoretische Physik, Universit\"at Regensburg, \\
D-93040 Regensburg, Germany}
\vspace{0.5in}
\begin{abstract}

Recent developments showed that hadron light-cone parton distributions could be directly extracted from spacelike correlators, known as quasi parton distributions, in the large hadron momentum limit.
Unlike the normal light-cone parton distribution, a quasi parton distribution contains ultraviolet (UV) power divergence associated with the Wilson line self energy. 
We show that to all orders in the coupling expansion, the power divergence can be removed by a ``mass" counterterm in the auxiliary $z$-field formalism, in the same way as the renormalization of power divergence for an open Wilson line. After adding this counterterm, the quasi quark distribution is improved such that it contains at most logarithmic divergences. Based on a simple version of discretized gauge action, we present the one-loop matching kernel between the improved non-singlet quasi quark distribution with a lattice regulator and the corresponding quark distribution in dimensional regularization. 
\end{abstract}

\maketitle

\section{introduction}
One of the most important goals of QCD is to understand the hadron structure from its fundamental degrees of freedom - quarks and gluons. This necessarily goes to the nonperturbative regime of QCD, and it is difficult in general to obtain first principle results directly from the QCD Lagrangian. A powerful tool of obtaining such results is Lattice QCD, which is an approach defined on Euclidean spacetime, and has been used to calculate hadron masses, charges etc. to a remarkable accuracy. However, it cannot be used to directly access intrinsically Minkowskian quantities such as the parton distribution functions (PDFs). PDFs are defined as the forward hadronic matrix elements of light-cone correlations, and describe the momentum distribution of quarks and gluons inside the hadron. They play a crucial role in understanding the experimental data at high energy hadron colliders such as the LHC. Owing to their real time dependence, Lattice QCD can only be used to indirectly extract the information on the PDFs by calculating their moments, which is limited by technical complications such as operator mixing. Another commonly used approach to determine PDFs is to assume a suitable parametrized form and fit to a large variety of experimental data. Most PDF groups obtained their PDF sets in this way~\cite{Alekhin:2012ig,Gao:2013xoa,Radescu:2010zz,CooperSarkar:2011aa,Martin:2009iq,Ball:2012cx}. A main drawback of this approach is that it suffers from parametrization uncertainty, and different groups usually produce different results for the same PDF.

Recent developments~\cite{Ji:2013fga,Ji:2013dva,Xiong:2013bka,Lin:2014zya,Hatta:2013gta,Ma:2014jla,Ji:2014hxa,Ji:2014gla,Ji:2014lra,
Alexandrou:2015rja,Ji:2015jwa,Ji:2015qla,Xiong:2015nua,Li:2016amo,Chen:2016utp} showed that the light-cone observables such as the PDFs can be directly extracted from the large momentum limit of the hadronic matrix element of a spacelike correlator, which is known as the quasi observable, using a large momentum effective theory~\cite{Ji:2014lra} (for other approaches to extract light-cone quantities see e.g.~\cite{Davoudi:2012ya,Detmold:2005gg,Liu:1993cv,Liu:1998um,Liu:1999ak,Liu:2016djw,Braun:2007wv}). The quasi PDF does not have a real time dependence, and thus can be simulated on the lattice. The infrared (IR) behaviors between the flavor non-singlet quasi-PDF and its corresponding light-cone PDF are shown to be the same at one loop by direct computations, and argued to be the same at all loops in Ref.~\cite{Xiong:2013bka}, based on which a factorization formula was also presented.

The factorization in Ref.~\cite{Xiong:2013bka} was given for the bare quasi quark distribution, where all fields and couplings entering the quasi distribution are bare ones. However, as the light-cone distribution, the quasi distribution also contains ultraviolet (UV) divergences and therefore needs renormalization. Ref.~\cite{Ji:2015jwa} explores the renormalization property of the quasi distribution, and shows that it is multiplicatively renormalizable at two-loop order. Also an equivalence was established between the virtual correction of the quasi quark distribution and the correction to the heavy-light quark vector current in heavy quark effective theory, so that the UV divergences in the former can be renormalized as the renormalization of the latter. Dimensional regularization was used in Ref.~\cite{Ji:2015jwa}, and the linear divergence present in a cutoff or lattice regularization was ignored. In realistic lattice calculations, one needs to know how to deal with such power divergences. This is one goal of the present paper. We will show that the power divergence in the quasi quark distribution can be removed to all-loop orders by a mass counterterm, which is the same as the renormalization of an open Wilson line. After such a mass renormalization, the quasi quark distribution is improved such that it contains at most logarithmic divergences.

The rest of this paper is organized as follows. In Sec. 2, we discuss the renormalization of power divergences arising from the Wilson line self energy in the quasi quark distribution. In Sec. 3, we present a lattice perturbation theory matching for the quasi quark distribution. We then conclude in Sec. 4.

\section{improved quasi quark distribution through Wilson line renormalization}
Let us start by recalling the definition of the quasi quark distribution. For the unpolarized quark density, it is given as~\cite{Ji:2013dva}
\beq\label{qUnpolDef}
   \tilde q(x, \Lambda, p^z) = \int^\infty_{-\infty} \frac{dz}{4\pi} e^{izk^z}  \langle p|\overline{\psi}(0, 0_\perp, z)
   \gamma^z L(z,0)\psi(0) |p\rangle \ ,
\eeq
where the quark fields are separated along the spatial $z$-direction, $L(z,0)$ is the Wilson line gauge link inserted to ensure gauge invariance, and $\Lambda$ denotes the UV cutoff.

\begin{figure}[tbp]
\centering
\includegraphics[width=0.2\textwidth]{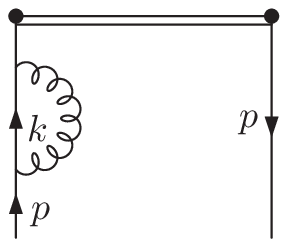}
\hspace{2em}
\includegraphics[width=0.2\textwidth]{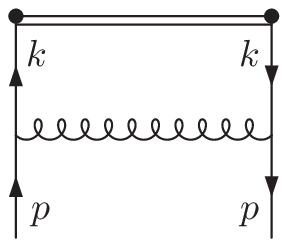}
\hspace{2em}
\includegraphics[width=0.2\textwidth]{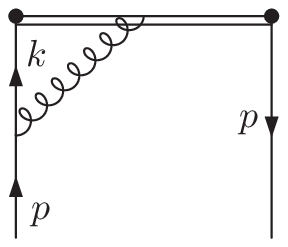}
\hspace{2em}
\includegraphics[width=0.2\textwidth]{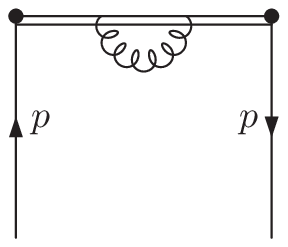}
\caption{One-loop diagrams for quasi quark distribution in Feynman gauge, the conjugate diagrams are not shown.}
\label{1loopFeyn}
\end{figure}

The one-loop correction to the above quark density has been computed both in the axial gauge~\cite{Xiong:2013bka} and in the Feynman gauge~\cite{Ji:2015jwa}. In Ref.~\cite{Ji:2015jwa}, the Wilson line $L(z,0)$ was separated into two Wilson lines along the $z$-axis $L(\infty,0)$ and $L(z,\infty)$, and associated respectively to the quark field to form gauge invariant combinations. Here we keep the gauge link as $L(z,0)$. The one-loop diagrams are then given in Fig.~\ref{1loopFeyn}. It is straightforward to check that these diagrams yield the same one-loop result as that in Ref.~\cite{Ji:2015jwa}. The linear divergence comes from the last diagram, which is a Wilson line self energy. It is known that such a linear divergence can be removed by a mass renormalization with the help of an auxiliary $z$-field formalism~\cite{Dotsenko:1979wb,Dorn:1986dt,Craigie:1980qs}, where the $z$-field is defined in a one-dimensional parameter space, and the non-local Wilson line can be interpreted as a two-point function of the $z$-field. In a sense, the auxiliary field $Z(z)$ can be understood as a Wilson line extending between $[z,\infty]$, i.e.
\beq
Z(z)=L(z,\infty),
\eeq
which satisfies an equation of motion 
\beq
\left[\partial_z-i g A_z(z)\right]Z(z)=0 ,
\eeq
analogous to a heavy quark field (in the heavy quark limit).   
The Wilson line 
\beq
L(z,0)=Z(z)Z^{\dagger}(0) 
\eeq
also renormalizes analogously to a heavy quark two point function 
\beq\label{WLrenorm}
L^{\text{ren}}(z,0)=\mathcal{Z}_Z^{-1}e^{-\delta m |z|}L(z,0) ,
\eeq
where $\delta m$ is analogous to the dimension one heavy quark mass counterterm and is linearly divergent, while the dimension zero wave function renormalization constant $\mathcal{Z}_Z$ is logarithmically divergent~\cite{Dotsenko:1979wb}.

To illustrate the cancellation of linear divergence from the mass counterterm, let us determine $\delta m$ at one-loop level. Recall that the last diagram in Fig.~\ref{1loopFeyn} leads to the following linear divergence
\beq
\lim_{\epsilon\to 0}\int dk_z\frac{\alpha_s C_F\Lambda}{2\pi}\frac{[\delta(k_z-\bar x p_z)-\delta(\bar x p_z)]p_z}{k_z^2+\epsilon^2}
\eeq
with $\bar x=1-x$. On the other hand, plugging Eq.~(\ref{WLrenorm}) into Eq.~(\ref{qUnpolDef}), one finds that the mass counterterm yields the following contribution
\begin{align}
&-\int \frac{dz}{2\pi}p_z\, e^{i(x-1)p_z z}|z|\, \delta m=-\lim_{\epsilon\to 0}\int \frac{dz}{2\pi}p_z\, e^{-i \bar x p_z z}\frac{1-e^{-\epsilon |z|}}{\epsilon}\delta m\non\\
&=-\lim_{\epsilon\to 0}\int \frac{dz}{2\pi}p_z\, e^{-i \bar x p_z z}\int_0^\infty \frac{d\alpha}{\sqrt{\pi \alpha}}(1-e^{-\frac{z^2}{4\alpha}})e^{-\alpha \epsilon^2}\delta m\non\\
&=-\lim_{\epsilon\to 0}\int \frac{dz}{2\pi}p_z\, e^{-i \bar x p_z z}\int_0^\infty d\alpha \int d k_z \frac{e^{-\alpha (k_z^2+\epsilon^2)}(1-e^{i k_z z})}{\pi} \delta m\non\\
&=-\lim_{\epsilon\to 0}\int \frac{dz}{2\pi}p_z\, e^{-i \bar x p_z z}\int dk_z \frac{1}{\pi}\frac{(1-e^{i k_z z})}{k_z^2+\epsilon^2}\delta m\non\\
&=-\lim_{\epsilon\to 0}\int \frac{dk_z}{\pi}p_z \frac{\delta(\bar x p_z)-\delta(k_z-\bar x p_z)}{k_z^2+\epsilon^2}\delta m.
\end{align}
We therefore have
\beq
\delta m=-\frac{\alpha_s C_F}{2\pi}(\pi\Lambda)
\eeq
at one-loop. 

In coordinate space, by expanding the Wilson line to $\mathcal O(g^2)$, we have
\beq\label{WLcoordspace}
g^2 C_F\int_0^z dz_1\int_0^{z_1} dz_2 \frac{1}{4\pi^2}\frac{1}{(z_1-z_2)^2+a^2}=\frac{g^2 C_F}{4\pi^2}\big[\frac{z}{a}\tan^{-1}\big(\frac z a\big)-\frac 1 2\ln\big({1+\frac{z^2}{a^2}}\big)\big],
\eeq
where we have introduced a cutoff $a$ to regularize the short distance singularity in the coordinate space gluon propagator when $z_1\to z_2$. By Fourier transforming to momentum space, the above result leads to
\beq\label{eq8}
\frac{g^2 C_F}{8\pi^2}\big[\frac{1}{a p_z(1-x)^2}-\frac{1}{|1-x|}+C\delta(1-x)\big],
\eeq
where $C$ is a constant that can be determined by demanding the $x$-integration of the above expression vanish. We can therefore identify
\beq
\Lambda=\frac{1}{a}.
\eeq

\begin{figure}[tbp]
\centering
\includegraphics[width=0.2\textwidth]{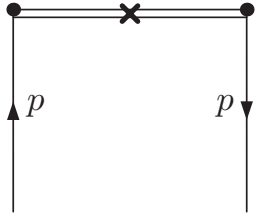}
\caption{One-loop mass counterterm diagram for quasi quark distribution in Feynman gauge.}
\label{1loopFeynmassCT}
\end{figure}

The mass counterterm diagram of order $\alpha_s$ is shown in Fig.~\ref{1loopFeynmassCT}. The gauge independence of $\delta m$ can already be seen from that the one-loop correction in the axial gauge\B{~\cite{Xiong:2013bka}}  and Feynman gauge\B{~\cite{Ji:2015jwa}} contains identical linear divergences. By using the $R_\xi$ gauge, one easily finds that the extra piece $k^\mu k^\nu/k^2$ (where $k$ is the gluon momentum) in the numerator of gluon propagator has no impact on the linear divergence, therefore $\delta m$ remains the same in different gauges.

One can add more gluons and quarks to the one-loop diagrams in Fig.~\ref{1loopFeyn} to form higher loop diagrams. Power counting tells us that power divergence appears only in those diagrams which contain the Wilson line self energy as a subdiagram. As shown in the appendix, beyond one-loop the mass counterterm also removes all the power divergence from the Wilson line. Therefore, after including the mass counterterm contribution, there is at most logarithmic divergence in the quasi quark distribution. This indicates that, as far as the power divergence is concerned, the renormalization of a quark bilinear operator like the quasi quark distribution is the same as the renormalization of an open Wilson line. In the discussions above, we showed that the linear divergence is indeed cancelled by the mass counterterm at perturbative one-loop level. On the lattice, the mass counterterm $\delta m$ has to be determined nonperturbatively, which can be done e.g. as in Ref.~\cite{Musch:2010ka}.

\section{matching between lattice and continuum through lattice perturbation theory}
\begin{figure}[tbp]
\centering
\includegraphics[width=0.2\textwidth]{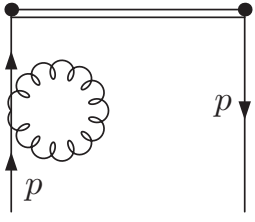}
\caption{The extra one-loop diagram for quasi quark distribution on the lattice, the conjugate diagram is not shown.}
\label{1loopLQCD}
\end{figure}

In this section, we aim at establishing the one-loop matching connecting the quasi quark distribution on the lattice to the normal quark distribution in the continuum. On the lattice, besides the diagrams in Fig.~\ref{1loopFeyn}, there exists an extra one-loop diagram (shown in Fig.~\ref{1loopLQCD}). 

The first diagram in Fig.~\ref{1loopFeyn} and the diagram in Fig.~3 are nothing but the Feynman gauge quark field renormalization, and have been well understood in lattice QCD. Their contribution can also be obtained from the other diagrams in Fig.~\ref{1loopFeyn} through quark number conservation. As shown in Refs.~\cite{Xiong:2013bka} and~\cite{Ji:2015jwa}, in the continuum the second and third diagrams in Fig.~\ref{1loopFeyn} contain a logarithmic dependence on the hadron momentum instead of a logarithmic divergence. Their contribution is independent of the regularization and thus will be the same in the continuum and on the lattice. One only needs to  match the linearly divergent part of the last diagram in Fig.~\ref{1loopFeyn} between lattice and continuum. This gives a connection between the lattice cutoff and the linear divergence in the one-loop kernel in the continuum~\cite{Xiong:2013bka} (in general they will differ by a finite term). 
To obtain this connection, we compute the Wilson line self energy using lattice perturbation theory based on a simple version of discretized gauge action.

The contribution of the Wilson line self energy diagram can be written as
\begin{align}
\delta W&=\frac{\alpha_s C_F}{8\pi^3}\int d^4(k a_L) \frac{1-\cos(k_z z)}{(1-\cos(k_z a_L))4\sum_\lambda \sin^2\frac{k_\lambda a_L}{2}}\non\\
&=\alpha_s C_F[\frac{\pi}{2}\frac{|z|}{a_L}-\frac{1}{\pi}\ln\frac{\pi |z|}{a_L}-\frac{1}{\pi}]+\mathcal O(\frac{a_L}{z}),
\end{align}
where $a_L$ denotes the lattice spacing. Comparing this with the expansion of Eq.~(\ref{WLcoordspace}) leads to the following matching between the linear divergence in the continuum and on the lattice
\beq
\Lambda=\frac{\pi}{a_L}.
\eeq
By replacing $\Lambda$ in the one-loop kernel of Ref.~\cite{Xiong:2013bka} with the above matching condition, we then obtain the one-loop kernel relating the quasi PDF on the lattice to the normal PDF. By adding a factor of $e^{-\delta m |z|}$ to the integrand of Eq.~(\ref{qUnpolDef}), we obtain the improved quasi-PDF $\tilde q_{\text{imp}}$ free of power divergence
\beq\label{im}
   \tilde q_{\text{imp}}(x, \Lambda, p^z) = \int^\infty_{-\infty} \frac{dz}{4\pi} e^{izk^z-\delta m |z|}  \langle p|\overline{\psi}(0, 0_\perp, z)
   \gamma^z L(z,0)\psi(0) |p\rangle \ .
\eeq

The matching kernel $Z$ between the improved quasi-PDF $\tilde q_{\text{imp}}$ and the PDF $q$ is defined as
\beq
\tilde q_{\text{imp}}(x, a_L, p^z)  = \int^1_{-1} \frac{dy}{|y|} Z\left(\frac{x}{y},{p^z a_L},\frac{\mu}{p^z}\right) q(y, \mu) +  {\cal O}\left(\Lambda^2_{\rm QCD}/(p^z)^2,  M^2/(p^z)^2\right)\ .
\eeq
At one-loop the $Z$ factor can be extracted from Eqs.~(15)-(19) of Ref.~\cite{Xiong:2013bka} with the $1/a_L$ contribution 
subtracted by the counterterm diagram in Fig.~\ref{1loopFeynmassCT}.
\begin{equation}
             Z\left(\xi,  p^z a_L,\frac{\mu}{p^z}\right) = \delta (\xi-1) + \frac{\alpha_s}{2\pi} Z^{(1)}
             \left(\xi, p^z a_L,\frac{\mu}{p^z}\right) + \dots ,
\end{equation}
where
\begin{align}
Z^{(1)}/C_F
=\left\{ \begin{array} {ll} \left(\frac{1+\xi^2}{1-\xi}\right)\ln \frac{\xi}{\xi-1} + 1\ , & \xi>1\ , \\ \left(\frac{1+\xi^2}{1-\xi}\right)\ln \frac{(p^z)^2}{\mu^2}+\left(\frac{1+\xi^2}{1-\xi}\right)\ln\big[{4\xi(1-\xi)}\big]-\frac{2\xi}{1-\xi}+1\ , & 0<\xi<1\ , \\ \left(\frac{1+\xi^2}{1-\xi}\right)\ln \frac{\xi-1}{\xi}-1\ , & \xi<0\ , \end{array} \right.
\end{align}
and the extra contribution near $\xi=1$ is
\begin{align}
\delta Z^{(1)}
=\frac{\alpha_S C_F}{2\pi} \int dy \left\{ \begin{array} {ll} -\frac{1+y^2}{1-y}\ln \frac{y}{y-1}-1\ , & y>1\ , \\ -\frac{1+y^2}{1-y}\ln \frac{(p^z)^2}{\mu^2}-\frac{1+y^2}{1-y}\ln\big[4y(1-y)\big]+\frac{2y(2y-1)}{1-y}+1\ , & 0<y<1\ , \\ -\frac{1+y^2}{1-y}\ln \frac{y-1}{y}+1\ , & y<0\ . \end{array} \right.
\end{align}
Given that the power divergence has been removed, the above one-loop matching kernel facilitates a reliable extraction of normal PDFs from lattice data.

\section{Conclusion}
In this paper, we have shown that the power divergence present in the lattice regularization of the quasi PDFs can be removed to all-loop orders by a mass counterterm, which can be interpreted as the mass renormalization for a test particle moving on the Wilson line. After such a mass renormalization, the quasi distribution is improved such that it contains at most logarithmic divergences. We also present the one-loop matching for the quasi quark distribution between lattice and continuum using lattice perturbation theory based on a simple version of discretized gauge action. Our results facilitate a reliable extraction of physical PDFs from lattice data.

\vspace{1em}
We thank Will Detmold, David Lin and Iain Stewart for useful discussions. This work was partially supported by the U.S. Department of Energy via grants DE-FG02-93ER-
40762, a grant (No. 11DZ2260700) from the Office of Science and Technology in Shanghai Municipal Government,
grants from National Science Foundation of China (No. 11175114, No. 11405104), a DFG grant SCHA 458/20-1, the
Ministry of Science and Technology, Taiwan under Grant Nos. 105-2112-M-002 -017 -MY3 and 105-2918-I-002 -003
and the CASTS of NTU.

\vspace{2em}
\begin{center}
\bf{Note added}
\end{center}

While this work is being finalized, a preprint by Jian-Wei Qiu et al.~\cite{Ishikawa:2016znu} dealing with the same topic has appeared, where they obtain similar conclusions as ours on the renormalization of power divergence. 


\section{Appendix}
In this appendix, we present the arguments that the mass counterterm removes the power divergence to all-loop orders.

\begin{figure}[tbp]
\centering
\includegraphics[width=0.2\textwidth]{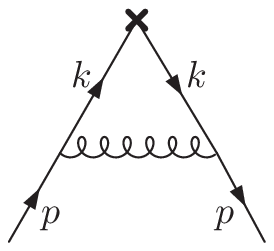}
\hspace{2em}
\includegraphics[width=0.2\textwidth]{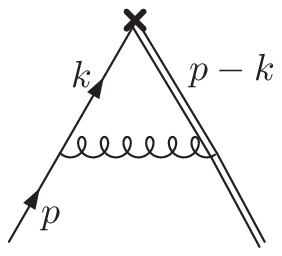}
\hspace{2em}
\includegraphics[width=0.2\textwidth]{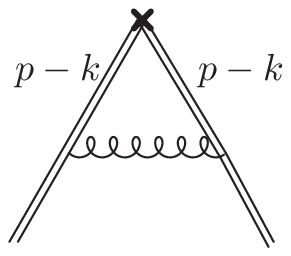}
\caption{One-loop diagrams to illustrate the real contribution of quasi quark distribution in Feynman gauge.}
\label{1loopFeynvi}
\end{figure}

For simplicity, we start with the one-loop diagrams for the quasi quark distribution in Ref.~\cite{Ji:2015jwa}, 
which can be equivalently drawn as the diagrams in Fig.~\ref{1loopFeynvi} with a vertex insertion (1 for virtual diagrams, and $\gamma^z\delta(x-k^z/p^z)$ for real diagrams to impose momentum constraint). At one-loop, it has been shown~\cite{Ji:2015jwa} that the virtual diagrams are respectively equivalent to the wave function renormalization of a light quark field, the renormalization of a heavy-light vector current, and the wave function renormalization of a heavy quark field. All these can be done to all-loop orders. Moreover, the mass counterterm can be chosen to exactly cancel the linearly divergent contribution from the third diagram in Fig.~\ref{1loopFeynvi} (both virtual contribution with vertex insertion 1 and real contribution with $\gamma^z\delta(x-k^z/p^z)$). Following Ref.~\cite{Dorn:1986dt}, this also holds to all-loop orders. After this mass renormalization, as the real diagrams contain an extra $\delta$-function which effectively reduces the UV degree of divergence by one, they will produce at most logarithmic divergences at higher-loop orders. Therefore, as far as power divergence is concerned, the mass renormalization is sufficient to remove them to all-loop orders.

\begin{thebibliography}{99}


\bibitem{Alekhin:2012ig}
  S.~Alekhin, J.~Blumlein and S.~Moch,
  Phys.\ Rev.\ D {\bf 86}, 054009 (2012)
  [arXiv:1202.2281 [hep-ph]].

\bibitem{Gao:2013xoa}
  J.~Gao, M.~Guzzi, J.~Huston, H.~-L.~Lai, Z.~Li, P.~Nadolsky, J.~Pumplin and D.~Stump {\it et al.},
  arXiv:1302.6246 [hep-ph].

\bibitem{Radescu:2010zz}
{\bf H1 and ZEUS} Collaborations, V.~Radescu, 
  {\em PoS} {\bf
  ICHEP2010} (2010) 168.

\bibitem{CooperSarkar:2011aa}
  A.~M.~Cooper-Sarkar [ZEUS and H1 Collaborations],
  PoS EPS {\bf -HEP2011}, 320 (2011)
  [arXiv:1112.2107 [hep-ph]].


\bibitem{Martin:2009iq}
  A.~D.~Martin, W.~J.~Stirling, R.~S.~Thorne and G.~Watt,
  Eur.\ Phys.\ J.\ C {\bf 63}, 189 (2009)
  [arXiv:0901.0002 [hep-ph]].

\bibitem{Ball:2012cx}
  R.~D.~Ball, V.~Bertone, S.~Carrazza, C.~S.~Deans, L.~Del Debbio, S.~Forte, A.~Guffanti and N.~P.~Hartland {\it et al.},
  Nucl.\ Phys.\ B {\bf 867}, 244 (2013)
  [arXiv:1207.1303 [hep-ph]].







\bibitem{Ji:2013fga}
  X.~Ji, J.~-H.~Zhang and Y.~Zhao,
  Phys.\ Rev.\ Lett.\  {\bf 111}, no. 11, 112002 (2013)
  [arXiv:1304.6708 [hep-ph]].


\bibitem{Ji:2013dva}
  X.~Ji,
  Phys.\ Rev.\ Lett.\  {\bf 110}, no. 26, 262002 (2013)
  [arXiv:1305.1539 [hep-ph]].


\bibitem{Xiong:2013bka} 
  X.~Xiong, X.~Ji, J.~H.~Zhang and Y.~Zhao,
  Phys.\ Rev.\ D {\bf 90}, 014051 (2014)
  [arXiv:1310.7471 [hep-ph]].


\bibitem{Lin:2014zya}
  H.~-W.~Lin, J.~-W.~Chen, S.~D.~Cohen and X.~Ji,
  arXiv:1402.1462 [hep-ph].


\bibitem{Hatta:2013gta} 
  Y.~Hatta, X.~Ji and Y.~Zhao,
  Phys.\ Rev.\ D {\bf 89}, 085030 (2014)
  [arXiv:1310.4263 [hep-ph]].

\bibitem{Ma:2014jla} 
  Y.~-Q.~Ma and J.~-W.~Qiu,
  arXiv:1404.6860 [hep-ph].

\bibitem{Ji:2014gla} 
  X.~Ji,
  Sci.\ China Phys.\ Mech.\ Astron.\  {\bf 57}, no. 7, 1407 (2014)
  [arXiv:1404.6680 [hep-ph]].

\bibitem{Ji:2014hxa} 
  X.~Ji, P.~Sun, X.~Xiong and F.~Yuan,
  arXiv:1405.7640 [hep-ph].

\bibitem{Ji:2014lra} 
  X.~Ji, J.~H.~Zhang and Y.~Zhao,
  arXiv:1409.6329 [hep-ph].

\bibitem{Alexandrou:2015rja} 
  C.~Alexandrou, K.~Cichy, V.~Drach, E.~Garcia-Ramos, K.~Hadjiyiannakou, K.~Jansen, F.~Steffens and C.~Wiese,
  Phys.\ Rev.\ D {\bf 92}, 014502 (2015)
  doi:10.1103/PhysRevD.92.014502
  [arXiv:1504.07455 [hep-lat]].

\bibitem{Ji:2015jwa} 
  X.~Ji and J.~H.~Zhang,
  Phys.\ Rev.\ D {\bf 92}, 034006 (2015)
  doi:10.1103/PhysRevD.92.034006
  [arXiv:1505.07699 [hep-ph]].

\bibitem{Ji:2015qla} 
  X.~Ji, A.~Schäfer, X.~Xiong and J.~H.~Zhang,
  Phys.\ Rev.\ D {\bf 92}, 014039 (2015)
  doi:10.1103/PhysRevD.92.014039
  [arXiv:1506.00248 [hep-ph]].

\bibitem{Xiong:2015nua} 
  X.~Xiong and J.~H.~Zhang,
  Phys.\ Rev.\ D {\bf 92}, no. 5, 054037 (2015)
  doi:10.1103/PhysRevD.92.054037
  [arXiv:1509.08016 [hep-ph]].

\bibitem{Li:2016amo} 
  H.~n.~Li,
  arXiv:1602.07575 [hep-ph].

\bibitem{Chen:2016utp} 
  J.~W.~Chen, S.~D.~Cohen, X.~Ji, H.~W.~Lin and J.~H.~Zhang,
  Nucl.\ Phys.\ B {\bf 911}, 246 (2016)
  doi:10.1016/j.nuclphysb.2016.07.033
  [arXiv:1603.06664 [hep-ph]].


\bibitem{Davoudi:2012ya} 
  Z.~Davoudi and M.~J.~Savage,
  Phys.\ Rev.\ D {\bf 86}, 054505 (2012)
  doi:10.1103/PhysRevD.86.054505
  [arXiv:1204.4146 [hep-lat]].

\bibitem{Detmold:2005gg} 
  W.~Detmold and C.~J.~D.~Lin,
  Phys.\ Rev.\ D {\bf 73}, 014501 (2006)
  doi:10.1103/PhysRevD.73.014501
  [hep-lat/0507007].

\bibitem{Liu:1993cv} 
  K.~F.~Liu and S.~J.~Dong,
  Phys.\ Rev.\ Lett.\  {\bf 72}, 1790 (1994)
  doi:10.1103/PhysRevLett.72.1790
  [hep-ph/9306299].

\bibitem{Liu:1998um} 
  K.~F.~Liu, S.~J.~Dong, T.~Draper, D.~Leinweber, J.~H.~Sloan, W.~Wilcox and R.~M.~Woloshyn,
  Phys.\ Rev.\ D {\bf 59}, 112001 (1999)
  doi:10.1103/PhysRevD.59.112001
  [hep-ph/9806491].

\bibitem{Liu:1999ak} 
  K.~F.~Liu,
  Phys.\ Rev.\ D {\bf 62}, 074501 (2000)
  doi:10.1103/PhysRevD.62.074501
  [hep-ph/9910306].

\bibitem{Liu:2016djw} 
  K.~F.~Liu,
  arXiv:1603.07352 [hep-ph].

\bibitem{Braun:2007wv} 
  V.~Braun and D.~Mueller,
  Eur.\ Phys.\ J.\ C {\bf 55}, 349 (2008)
  doi:10.1140/epjc/s10052-008-0608-4
  [arXiv:0709.1348 [hep-ph]].


\bibitem{Dotsenko:1979wb} 
  V.~S.~Dotsenko and S.~N.~Vergeles,
  Nucl.\ Phys.\ B {\bf 169}, 527 (1980).
  doi:10.1016/0550-3213(80)90103-0

\bibitem{Dorn:1986dt} 
  H.~Dorn,
  Fortsch.\ Phys.\  {\bf 34}, 11 (1986).
  doi:10.1002/prop.19860340104


\bibitem{Craigie:1980qs} 
  N.~S.~Craigie and H.~Dorn,
  Nucl.\ Phys.\ B {\bf 185}, 204 (1981).
  doi:10.1016/0550-3213(81)90372-2


\bibitem{Musch:2010ka} 
  B.~U.~Musch, P.~Hagler, J.~W.~Negele and A.~Schafer,
  Phys.\ Rev.\ D {\bf 83}, 094507 (2011)
  doi:10.1103/PhysRevD.83.094507
  [arXiv:1011.1213 [hep-lat]].



\bibitem{Ishikawa:2016znu} 
  T.~Ishikawa, Y.~Q.~Ma, J.~W.~Qiu and S.~Yoshida,
  arXiv:1609.02018 [hep-lat].



\end{thebibliography}
\end{document}